\documentclass[preprintnumbers,amsmath,amssymb,floatfix,10pt,prd,onecolumn,
superscriptaddress,nofootinbib]{revtex4}
\usepackage{latexsym}
\usepackage{epsfig}
\usepackage{epstopdf}
\usepackage{graphicx}
\usepackage{amssymb}
\usepackage{amsmath}
\usepackage{dcolumn}
\usepackage{bm}
\usepackage{color}
\usepackage{comment}

\begin{document}

\title{\bf Noncommutative inspired wormholes admitting conformal motion involving minimal coupling}

\author{M. Zubair}
\email{mzubairkk@gmail.com;drmzubair@cuilahore.edu.pk}\affiliation{Department
of Mathematics, COMSATS University Islamabad, Lahore Campus, Pakistan}

\author{Saira Waheed}
\email{swaheed@pmu.edu.sa}\affiliation{Prince Mohammad Bin Fahd University, Al Khobar, 31952 Kingdom of Saudi Arabia}

\author{G. Mustafa}
\email{gmustafa3828@gmail.com}\affiliation{Department of Mathematics, COMSATS University Islamabad, Lahore Campus, Pakistan}

\author{Hamood Ur Rehman}
\email{hamood84@gmail.com}\affiliation{Department of Mathematics, University of Okara, Okara, Pakistan}

\begin{abstract}

In this manuscript, we explore the existence of wormhole solutions
exhibiting spherical symmetry in a modified gravity namely $f(R,T)$
theory by involving some aspects of non-commutative geometry. For
this purpose, we consider the anisotropic matter contents along with
the well-known Gaussian and Lorentizian distributions of string
theory. For the sake of simplicity in analytic discussions, we take
a specific form of $f(R,T)$ function given by $f(R,T)=R+\lambda T$.
For both these non-commutative distributions, we get exact solutions
in terms of exponential and hypergeometric functions. By taking some
suitable choice of free parameters, we investigate different
interesting aspects of these wormhole solutions graphically. We also
explored the stability of these wormhole models using equilibrium
condition. It can be concluded that the obtained solutions are
stable and physically viable satisfying the wormhole existence
criteria. Lastly, we discuss the constraints for positivity of the
active gravitational mass for both these distributions.\\

\end{abstract}

\maketitle

\date{\today}

\section{Introduction}

One of the most interesting scientific outcomes of the previous
century is the accelerated expanding behavior of cosmos and its
responsible factor known as dark energy (DE) (an unknown nature of
energy density involving negative pressure). Although many
candidates are proposed for this unusual source, however, it still
remains as a matter of debate among the researchers that which
candidate could provide a successful explanation of its nature and
hence of the resultant rapid expansion of cosmos. In this regard,
the efforts can be grouped into two categories: modifications
adopted in matter sector of lagrangian and secondly, involvement of
some additional terms in gravity sector of action. Some important
members of the first group include tachyon model, quintessence,
Chaplygin gas and its different versions, phantom, quintom etc
\cite{1}. While, in the second approach, different modifications of
general relativity (GR) are proposed like telleparallel theory and
its generalized version $f(T)$ gravity, scalar-tensor gravity,
$f(T,T_G)$ with $T_G$ as Gauss-Bonnet alternative term, the $f(R)$
theory which is regraded as the basic generalization of GR obtained
by replacing the Ricci scalar with a generic function $f(R)$
\cite{2}.

In the construction of modified theories of gravity, a pioneer work
was presented by Harko et al. \cite{3} in 2014, where they proposed
a new kind of modification in $f(R)$ gravity by introducing an
interaction between Ricci scalar and matter sector, namely $f(R,T)$
theory. Later on, Houndjo et al. \cite{4} used this theory to
construct models generating accelerated cosmic expansion by taking a
special choice of $f(R,T)=f_1(R)+f_2(T)$ along with an auxiliary
scalar field. Further, in another study, they investigated $f(R,T)$
function numerically by taking holographic DE into account \cite{5}.
They concluded that their constructed function yields the same
stages of cosmic expansion as discussed in GR. In this respect,
Sharif and Zubair \cite{6} discussed the validity of thermodynamics
laws in the presence of holographic as well as new agegraphic DE in
this theory by reproducing $f(R,T)$ function. This theory is getting
more attention of the researchers recently and numerous interesting
aspects of this theory has been discussed in literature \cite{7}.

The tunnel or bridge type structures that provide a subway between
two different universes or two distant parts of the same universe
are referred as wormholes. In cosmology, the construction and
existence of wormholes are getting more attention of the researchers
day by day. Since wormhole requires exotic matter for their
existence, therefore in modified gravity theories, involving
modified energy-momentum tensor, this topic is regarded as one of
the most interesting issues under discussion. In GR, the
mathematical criteria for wormhole existence was presented by
Einstein and Rosen \cite{8} in $1938$ and their constructed
wormholes were labeled as Lorentzian wormholes or Schwarzchild
wormholes. In 1988, it was found \cite{9} that wormholes could be
large enough for humanoid travelers and even allow time travel.
Zubair et al. \cite{10} investigated the wormhole existence in
non-commutative $f(R,T)$ theory by taking two different models
$f_{1}(R)=R$ and $f_{1}(R)=R+\alpha R^{2}+\gamma R^{n}$ into
account. They found that the obtained wormholes solutions are
physically interesting and stable. In another study \cite{11}, they
discussed static spherically symmetric wormholes filled with
anisotropic, isotropic and barotropic fluids as three different
cases in $f(R,T)$ gravity. By considering Starobinsky $f(R)$ model,
they have shown that in few regions of spacetime, the wormhole
solutions can be discussed in the absence of exotic matter. In
different modifications of GR obtained by including some kind of
exotic matter like quintom, scalar field models, non-commutative
geometry and electromagnetic field etc., researchers have developed
different interesting and physically viable wormhole structures
\cite{12}.

The string theory and its well-known aspect of non-commutative
geometry is getting more attention of the researchers day by day.
The concept of non-commutativity emerges from the fact that the
coordinates may be treated as non-commutative operators on a
D-brane. This important property of string theory helps to
investigate mathematically some important concepts of quantum
gravity \cite{13}. Non-commutative geometry is basically an attempt
to unify the spacetime gravitational forces with weak and strong
forces on a single platform. In non-commutative geometry, one can
replace point-like structures by smeared objects and hence provides
spacetime discretization because of the commutator defined by the
relation $[x^{\alpha}, x^{\beta}] = i\theta^{\alpha\beta}$, where
$\theta^{\alpha\beta}$ denotes an anti-symmetric second-order
matrix. Gaussian distribution and Lorentizian distribution of
minimal length $\sqrt{\theta}$ can be used to model this smearing
effect instead of the Dirac delta function. The spherically
symmetric, static particle like gravitational source providing the
Gaussian distribution of non-commutative geometry with total mass
$M$ has energy density profile given by \cite{14}
\begin{equation}\label{n1}
\rho(r)=\frac{M}{(4\pi \theta)^{\frac{3}{2}}
}e^{-\frac{r^{2}}{4\theta}},
\end{equation}
while with reference to Lorentzian distribution, the density
function of particle-like mass $M$ can be written as follows
\begin{equation}\label{n2}
\rho(r)=\frac{M \sqrt{\theta}}{\pi^{2}(r^{2}+\theta)^{2}}.
\end{equation}
Here total mass $M$ can be considered as wormhole, a type of
diffused centralized object and clearly, $\theta$ is the
noncommutative parameter. In this respect, Sushkov \cite{15} has
used the Gaussian distribution source for modeling phantom-energy
upheld wormholes. Further, using this distribution, Nicolini and
Spalluci \cite{16} explained the physical impacts of
short-separation changes of non-commutative coordinates in the field
of black holes existence. Recently, Ghosh \cite{Ghosh} discussed the
Einstein-Gauss-Bonnet black holes in the background of
non-commutative geometry and they also presented thermodynamical
properties of the obtained solutions.

In this present paper, we investigate the spherically symmetric
wormhole existence by taking conformal killing vectors as well as
some important features of non-commutative geometry into account.
The present manuscript has been organized in this pattern. In the
next segment, we introduce $(R,T)$ gravity and its mathematical
formulation, i.e, field equations. In section \textbf{III}, a short
discussion on the conformal killing vectors for spherically
symmetric spacetime and the corresponding solutions will be given.
Also, we formulate the simplified form of field equations under the
light of conformal killing vectors there. In section \textbf{IV}, we
explore the existence of wormhole solutions by taking Gaussian and
Lorentzian distributions of non-commutative geometry mathematically
as well as graphically. Section \textbf{V} provides the stability of
the obtained solutions using equilibrium equation. Also, we explore
the criteria for the positivity of active gravitational mass there.
In the last section, we summarize the whole discussion by
highlighting major conclusions.

\section{Field Equations in $f(R,T)$ Gravity}

In 2014, Harko et al. \cite{3} presented a new generalization of
$f(R)$ gravity by taking the coupling of Ricci scalar with matter
field into account as follows
\begin{equation}\label{1}
S=\int \frac{f(R,T)}{16 \pi G}\sqrt{-g}d^{4}x+\int L_{m}\sqrt{-g}d^{4}x,
\end{equation}
where $f(R,T)$ is a generic function of $T$ and $R$ known as trace
of the energy momentum tensor $T_{\mu\nu}$ and Ricci scalar.
Further, $g_{\mu\nu}$ denotes the metric tensor while $L_{m}$ is the
matter Lagrangian density. This theory is considered to be more
successful as compared to $f(R)$ gravity in this sense that such a
theory can include quantum effects or imperfect fluids that are
neglected in a simple $f(R)$ generalization of GR. The variation of
above action with respect to metric tensor $g_{\mu\nu}$ yields the
following set of field equations:
\begin{eqnarray}\label{2}
8\pi
T_{\mu\nu}-f_{T}(R,T)T_{\mu\nu}-f_{T}(R,T)\Theta_{\mu\nu}&=&f_{R}(R,T)R_{\mu\nu}
-\frac{1}{2}f(R,T)g_{\mu\nu}+(g_{\mu\nu}\Box-\nabla_{\mu}\nabla_{\nu})f_{R}(R,T).
\end{eqnarray}
The contraction of the above equation leads to a relation between
Ricci scalar $R$ and trace $T$ of the energy momentum tensor as
follows:
\begin{eqnarray}\label{3}
8\pi T-f_{T}(R,T)T-f_{T}(R,T)\Theta=f_{R}(R,T)R+3\Box f_{R}(R,T)-2f(R,T).
\end{eqnarray}
These two equations involve covariant derivative and d'Alembert
operator denoted by $\nabla$ and $\Box$, respectively. Furthermore,
$f_{R}(R,T)$ and $f_{T}(R,T)$ correspond to the function derivatives
with respect to $R$ and $T$, respectively. Also, the term
$\Theta_{\mu\nu}$ is defined by
\begin{equation}\nonumber
\Theta_{\mu\nu}=\frac{g^{\alpha\beta}\delta T_{\mu\nu}}{\delta
g^{\mu\nu}}=-2T_{\mu\nu}+g_{\mu\nu}L_{m}-2g^{\alpha\beta}\frac{\partial^{2}L_{m}}{\partial
g^{\mu\nu}\partial g^{\alpha\beta}}.
\end{equation}
The anisotropic source of matter is defined by the energy-momentum
tensor given by
\begin{equation}\nonumber
T_{\mu\nu}=(\rho+p_{t})V_{\mu}V_{\nu}-p_{t}g_{\mu\nu}+(p_{r}-p_{t})\chi_{\mu}\chi_{\nu},
\end{equation}
where $V_{\mu}$ is the 4-velocity vector of the fluid given by
$V^{\mu}=e^{-a}\delta^{\mu}_{0}$ and $\chi^{\mu}=e^{-b}\delta^{\mu}_{1}$
which satisfy the relations: $V^{\mu}V_{\mu}=-\chi^{\mu}\chi_{\mu}=1$. Here
we choose $L_{m}=-\rho$, which leads to following expression for
$\Theta_{\mu\nu}$:
\begin{equation}\nonumber
\Theta_{\mu\nu}=-2T_{\mu\nu}-\rho g_{\mu\nu}.
\end{equation}
If we relate the trace equation (\ref{3}) with equation (\ref{2}),
then Einstein field equations can take the form given by
\begin{eqnarray}\nonumber
f_{R}(R,T)G_{\mu\nu}&=&(8\pi+f_{T}(R,T))T_{\mu\nu}+[\nabla_{\mu}\nabla_{\nu}f_{R}(R,T)\\\label{4}&-&
\frac{1}{4}g_{\mu\nu}\{(8\pi+f_{T}(R,T))T+\Box f_{R}(R,T)+f_{R}(R,T)R)\}].
\end{eqnarray}
The line element describing a static spherically symmetric geometry
can be written as
\begin{equation}\label{2.1}
ds^2=-e^{\mu(r)}dt^2+e^{\nu(r)}dr^2+r^{2}(d\theta^{2}+sin^{2}\theta
d\Phi^{2}),
\end{equation}
where $\mu(r)$ and $\nu(r)$ are the metric potentials dependent on
the radial coordinate $r$.

Here we are interested to find analytical wormhole solutions in the
background of non-commutative $f(R,T)$ gravity involving conformal
killing vectors. For this purpose, we choose  $f(R,T)=R+\lambda T$
to formulate the modified field equations (\ref{4}) with the
wormhole space-time (\ref{2.1}), the resulting expressions of energy
density, radial and transverse stresses are found to be
\begin{eqnarray}
\rho&=&\frac{e^{-\nu(r)}}{4 \left(2 \lambda ^2+1\right) r^2} \bigg[r \left(2 \lambda  r \mu''(r)+\lambda  \left(r \mu'(r)+4\right)
\left(\mu'(r)-\nu'(r)\right)+4 \nu'(r)\right)-4 (\lambda -1)
\left(e^{\nu(r)}-1\right)\bigg],\label{4.4}\\
p_{r}&=&\frac{e^{-\nu(r)}}{4 (\lambda +1) \left(2 \lambda ^2+1\right) r^2} \bigg[r \left(-2 (\lambda -1) \lambda  r \mu''(r)+\mu'(r)
\left((\lambda -1) \lambda  r \nu'(r)+4 \left(\lambda ^2+\lambda
+1\right)\right)\right.+\left(\lambda -1) \lambda  (-r) \mu'(r)^2\right.\nonumber\\&-&\left.4 \lambda  (\lambda +2)
\nu'(r)\right)-4 (\lambda  (3 \lambda +2)+1) \left(e^{\nu(r)}-1\right)\bigg],\label{4.5}\\
p_{t}&=&\frac{e^{-\nu(r)}}{4 (\lambda +1) \left(2 \lambda ^2+1\right) r^2} r \bigg[2 \left(\lambda ^2+\lambda +1\right) r
\mu''(r)+\mu'(r) \left(-\left(\lambda ^2+\lambda +1\right)\right)\bigg]\times\bigg[\left( r \nu'(r)+4 \lambda
+2\right)\nonumber \\ &+&\left(\lambda ^2+\lambda +1\right) r \mu'(r)^2 -2 (2 \lambda +1)^2 \nu'(r)\bigg]-4 \lambda  (\lambda +2)
\left(e^{\nu(r)}-1\right).\label{4.6}
\end{eqnarray}
Here clearly, in the limit $\lambda=0$, the field equations of GR
can be recovered.

\section{Wormhole geometries admitting conformal killing vectors}

In general, conformal Killing vectors (CKVs) explain the
mathematical relation between the geometry and contents of matter in
the spacetime via Einstein set of field equations. The CKVs are used
to generate the exact solution of Einstein field equation in more
convenient form as compared to other analytical approaches. Further,
these are used to discover the conservation laws in any spacetime.
The Einstein field equations being highly non-linear partial
differential equations can be reduced to a set of ordinary
differential equations by using CKVs.

Now we discuss the CKVs for spherically symmetric line element
(\ref{2.1}) and the corresponding field equations of $f(R,T)$
gravity. The conformal Killing vector is defined through the
relation
\begin{equation}\label{4.1}
\mathcal{L}_{\xi}g_{\mu\nu}=g_{\eta\nu}\xi^{\eta}_{;\mu}+g_{\mu\eta}\xi^{\eta}_{;\nu}=\psi(r)g_{\mu\nu},
\end{equation}
where $\mathcal{L}$ represents the Lie derivative of metric tensor
and $\psi(r)$ is the conformal vector. Using Eq.(\ref{2.1}) in
Eq.(\ref{4.1}), we get the following relations:
\begin{eqnarray*}
\xi^{1}\mu^{'}(r)&=&\psi(r),\\
\xi^{1}&=&\frac{r\psi(r)}{2},\\
\xi^{1}\nu^{'}(r)+2\xi^{1}_{,1}&=&\psi(r),
\end{eqnarray*}
where prime denotes the derivatives with respect to radial
coordinates $r$. Integration of these equations imply
\begin{eqnarray}
e^{\mu(r)}&=&C_{1}^{2}r^2,\label{4.2}\\
e^{\nu(r)}&=&\left(\frac{C_{2}}{\psi}\right)^{2},\label{4.3}
\end{eqnarray}
where $C_1$ and $C_2$ are constants of integration.

Using Eqs.(\ref{4.2}) and (\ref{4.3}) in
Eqs.(\ref{4.4})-(\ref{4.6}), we have the following expressions of
density, radial as well as tangential pressures:
\begin{eqnarray}
\rho&=&\frac{-2 C_2^2 (\lambda -1)+(6 \lambda -2) \psi ^2(r)+2 (3 \lambda -2) r \psi (r) \psi '(r)}{2 C_2^2 \left(2 \lambda ^2+1\right)
r^2}, \;\;\;\;\;\;\;\;\label{4.7}\\
p_{r}&=&\frac{-2 C_2^2 (\lambda  (3 \lambda +2)+1)+2 (\lambda  (5 \lambda +4)+3) \psi ^2(r)+\lambda  (\lambda +5) r \psi (r) \psi
'(r)}{2
C_2^2 (\lambda +1) \left(2 \lambda ^2+1\right) r^2},\;\;\;\;\;\;\;\;\; \label{4.8}\\
p_{t}&=&\frac{-2 C_2^2 \lambda  (\lambda +2)+2 (\lambda  (\lambda +4)+1) \psi ^2(r)+(5 \lambda  (\lambda +1)+2) r \psi (r) \psi '(r)}{2
C_2^2 (\lambda +1) \left(2 \lambda ^2+1\right) r^2}. \;\;\;\;\;\;\;\;\label{4.9}
\end{eqnarray}
To solve the above system for $\psi(r)$, we have two possibilities:
one can either choose some specific form of $\rho$ or a relation
between $p_r$ and $p_t$. Here we prefer to pick $\rho$ in
noncommutative framework of string theory.

\section{Wormholes Existence in Gaussian and Lorentzian Distributed Noncommutative Frameworks}

In this section, we explore the existence of wormhole solutions in
the presence of Gaussian and Lorentzian distributions of string
theory. Also we will discuss the wormhole properties using graphical
approach.

\subsection{Wormhole Existence with Gaussian Distribution}

In view of essential aspects of non-commutativity approach which is
specifically sensitive to the Gaussian distribution of minimal
length $\sqrt{\theta}$, we utilize the mass density of a static,
spherically symmetric, smeared, particle-like gravitational source
given by (\ref{n1}). Comparing Eqs.(\ref{n1}) and (\ref{4.7}), we
get the following differential equation:
\begin{equation}\label{4.10}
\frac{-2 C_2^2 (\lambda -1)+(6 \lambda -2) \psi ^2(r)+2 (3 \lambda
-2) r \psi (r) \psi '(r)}{2 C_2^2 \left(2 \lambda
^2+1\right)r^2}=\frac{M}{(4\pi\theta)^{\frac{3}{2}}}e^{-\frac{r^{2}}{4\theta}}.
\end{equation}
Solving the above Eq.(\ref{4.10}), we find the relation for density
given by
\begin{eqnarray}
\psi^{2}(r)&=&\frac{1}{\pi ^{3/2} \sqrt{\theta } (3 \lambda -2) (3 \lambda -1)}C_{2}^{2} r^{\frac{3 \lambda }{3 \lambda -2}+\frac{2-6
\lambda }{3 \lambda -2}+1} \left(\frac{r^2}{\theta }\right)^{\frac{1}{3 \lambda -2}}\times \bigg[\pi ^{3/2} \sqrt{\theta }
\left(3 \lambda ^2-5 \lambda +2\right) \left(\frac{r^2}{\theta }\right)^{\frac{1}{2-3 \lambda }} \nonumber \\ &-&2^{\frac{3 \lambda }{3
\lambda -2}}
\left(6 \lambda ^3-2 \lambda ^2+3 \lambda -1\right)\bigg]\times\bigg[ M \left(\frac{r^2}{\theta }\right)^{\frac{3 \lambda
}{2-3 \lambda }} \Gamma \left(\frac{3-6 \lambda }{2-3 \lambda },\frac{r^2}{4 \theta }\right)\bigg]+D_{1} r^{\frac{2-6 \lambda }{3
\lambda -2}},\label{4.11}
\end{eqnarray}
where $D_{1}$ is an integration constants. Using this relation of
$\psi^{2}(r)$ in Eqs.(\ref{4.8}) and (\ref{4.9}), we get the
analytical forms of radial and tangential pressures as follows
\begin{eqnarray}
p_{r}&=&\frac{1}{8 (\lambda +1)}\Bigg[\frac{48 D_{1} (\lambda -1) r^{\frac{2}{2-3 \lambda }-4}}{C_{2}^{2} (3 \lambda -2)}-\frac{6
(\lambda -1) \left(2 \lambda ^2+1\right) M E_{\frac{1}{2-3 \lambda }-1}\left(\frac{r^2}{4 \theta }\right)}{\pi ^{3/2} \theta ^{3/2} (2-3
\lambda )^2}+\frac{\lambda  (\lambda +5) M e^{-\frac{r^2}{4 \theta }}}{\pi ^{3/2} \theta ^{3/2} (3 \lambda
-2)}\nonumber\\&+&\frac{8 (\lambda -1) (\lambda  (5 \lambda +4)+3)}{(3 \lambda -1) \left(2 \lambda ^2+1\right) r^2}-\frac{8 (\lambda  (3
\lambda
+2)+1)}{\left(2 \lambda ^2+1\right) r^2}\Bigg],\label{4.12}\\
p_{t}&=&\frac{}{8 (\lambda +1)}\Bigg[-\frac{48 D_{1} \lambda  r^{\frac{2}{2-3 \lambda }-4}}{C_{2}^{2} (3 \lambda -2)}+\frac{6 \lambda
\left(2 \lambda ^2+1\right) M E_{\frac{1}{2-3 \lambda }-1}\left(\frac{r^2}{4 \theta }\right)}{\pi ^{3/2} \theta ^{3/2} (2-3 \lambda
)^2}+\frac{(5 \lambda  (\lambda +1)+2) M e^{-\frac{r^2}{4 \theta }}}{\pi ^{3/2} \theta ^{3/2} (3 \lambda
-2)}\nonumber\\&+&\frac{8 (\lambda -1) (\lambda  (\lambda +4)+1)}{(3 \lambda -1) \left(2 \lambda ^2+1\right) r^2}-\frac{8 \lambda
(\lambda
+2)}{\left(2 \lambda ^2+1\right) r^2}\Bigg],\label{4.13}
\end{eqnarray}
where $E$ is an exponential integral function which is defined as
\begin{equation}
E_{n}(x)=-\int^{\infty}_{-x}\frac{e^{-t}}{t}dt.
\end{equation}
Now we define the metric potentials in the scope of redshift and
shape function as follows
\begin{eqnarray}
e^{\mu(r)}=e^{2\Phi(r)}, \quad e^{\nu(r)}=\frac{1}{1-\frac{b(r)}{r}}.
\end{eqnarray}
Therefore, redshift function and shape function are given by
\begin{eqnarray}\label{4.14}
\Phi(r)&=&\ln(c_{2}r),\\\label{4.15}
b(r)&=&\frac{-r}{C_{2}^{2}}\Bigg[r^{-\frac{2}{3 \lambda -2}-2} \bigg[-\frac{C_{2}^{2} \left(2 \lambda ^2+1\right) M r^{\frac{2}{3 \lambda
-2}+4} E_{\frac{1}{2-3 \lambda }-1}\left(\frac{r^2}{4 \theta }\right)}{8 \pi ^{3/2} \theta ^{3/2} (3 \lambda
-2)}+\frac{C_{2}^{2} (\lambda -1) r^{\frac{2}{3 \lambda -2}+2}}{3 \lambda
-1}+D_{1}\bigg]\Bigg]+\frac{r}{C_{2}^{2}}.
\end{eqnarray}
\begin{figure}
\centering \epsfig{file=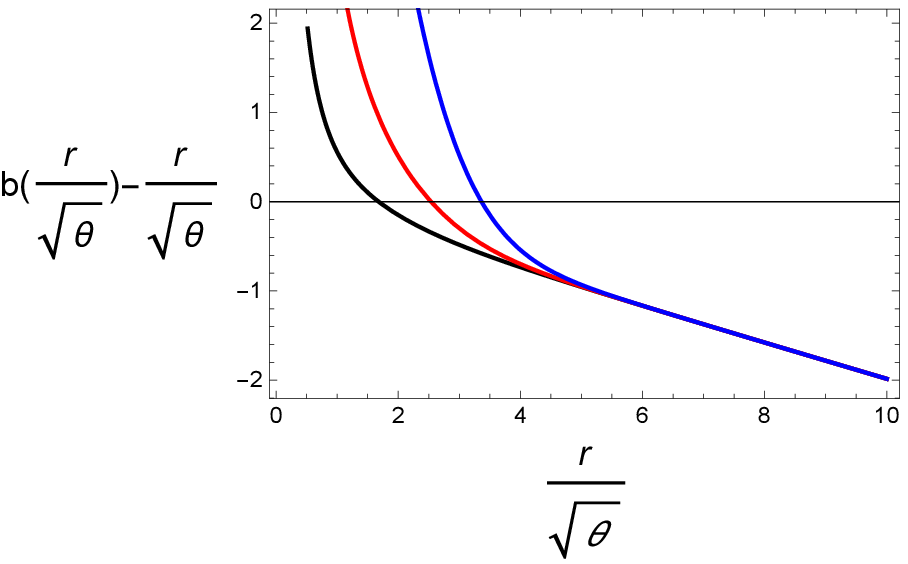, width=.45\linewidth,
height=1.4in}\epsfig{file=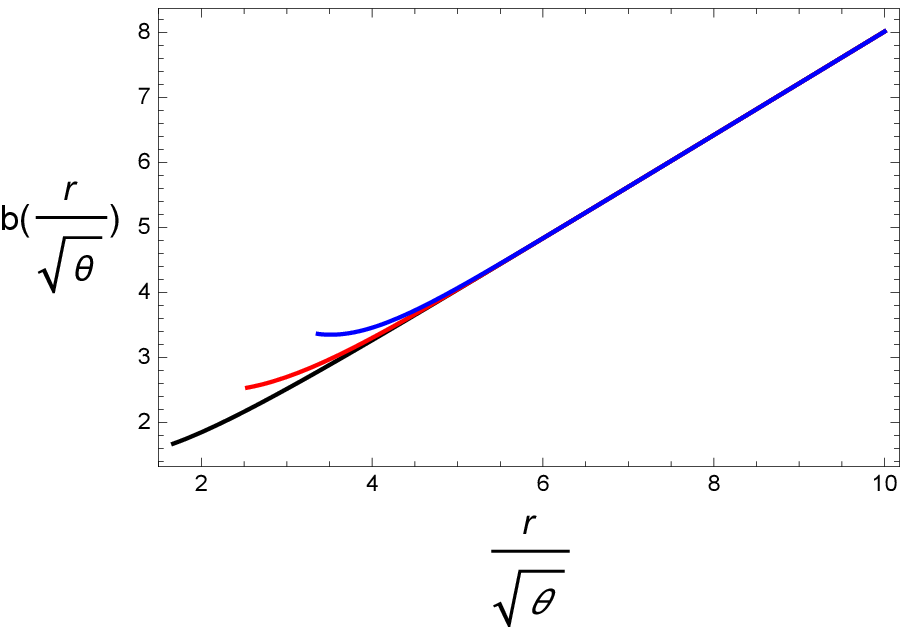, width=.45\linewidth,
height=1.4in}
\caption{\label{fig 4.1} Evolution of $b(\frac{r}{\sqrt{\theta}})-(\frac{r}{\sqrt{\theta}})$ and $b(\frac{r}{\sqrt{\theta}})$ versus $\frac{r}{\sqrt{\theta}}$ for different values of $\frac{M}{\sqrt{\theta}}$. Herein for Gaussian distribution, $\frac{M}{\sqrt{\theta}}=0.2$ represents the black curve, $\frac{M}{\sqrt{\theta}}=2$ represents the red curve and $\frac{M}{\sqrt{\theta}}=10$ represents the blue curve.}
\end{figure}
\begin{figure}
\centering \epsfig{file=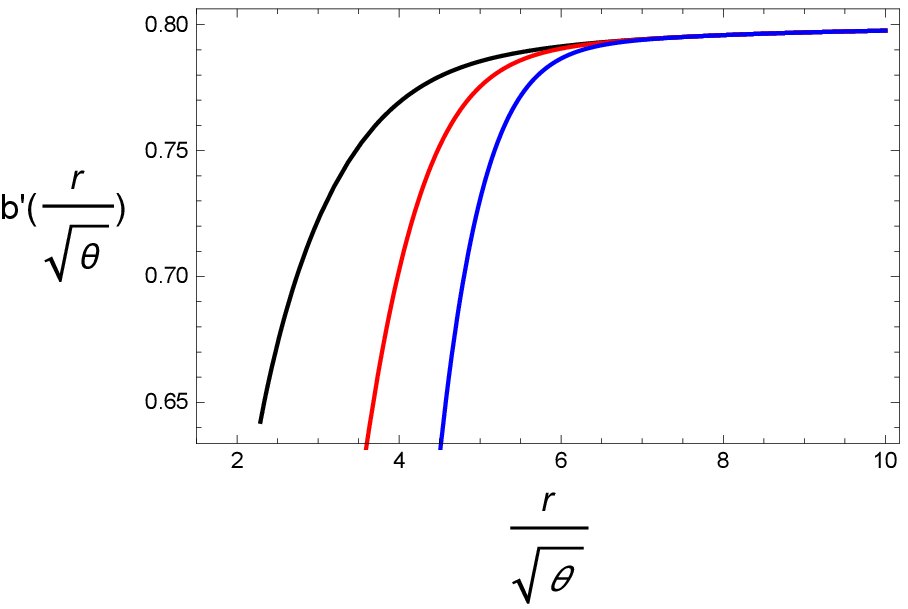, width=.45\linewidth,
height=1.4in}\epsfig{file=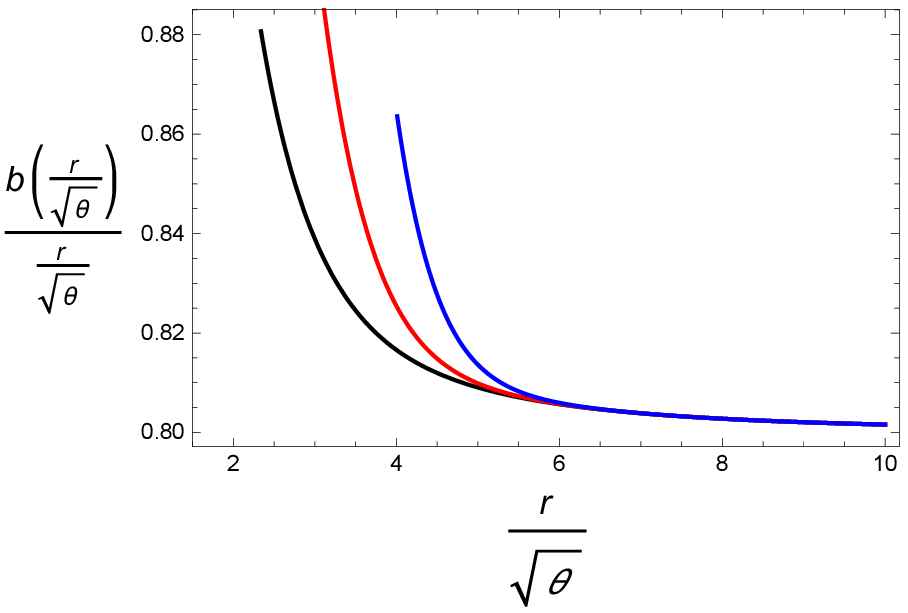, width=.45\linewidth,
height=1.4in}
\caption{\label{fig 4.2} This indicates the behavior of $b'(\frac{r}{\sqrt{\theta}})$ $\frac{b(\frac{r}{\sqrt{\theta}})}{\frac{r}{\sqrt{\theta}}}$ and versus $\frac{r}{\sqrt{\theta}}$ for different values of $\frac{M}{\sqrt{\theta}}$ in the framework of Gaussian distribution.}
\end{figure}
\begin{figure}
\centering
\epsfig{file=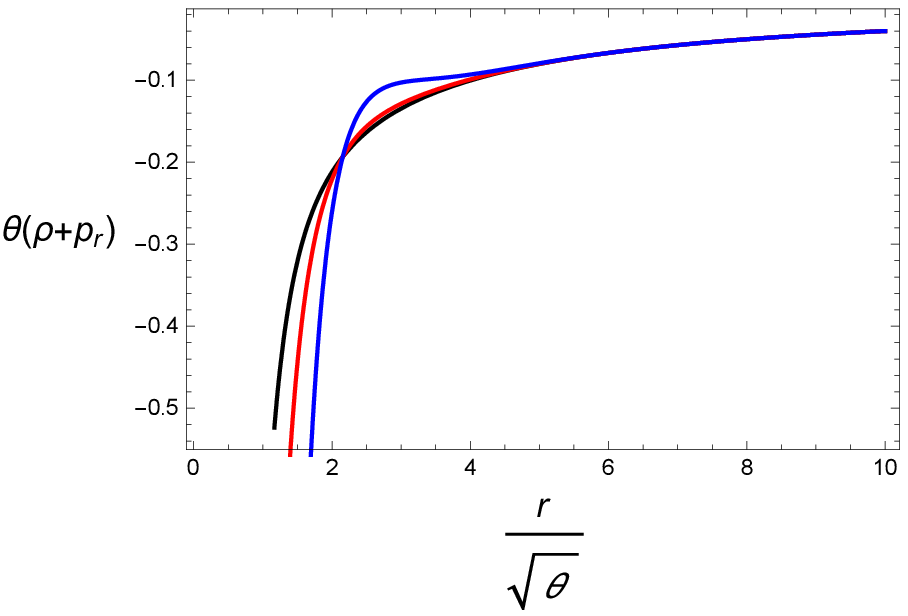, width=.45\linewidth,
height=1.4in}\epsfig{file=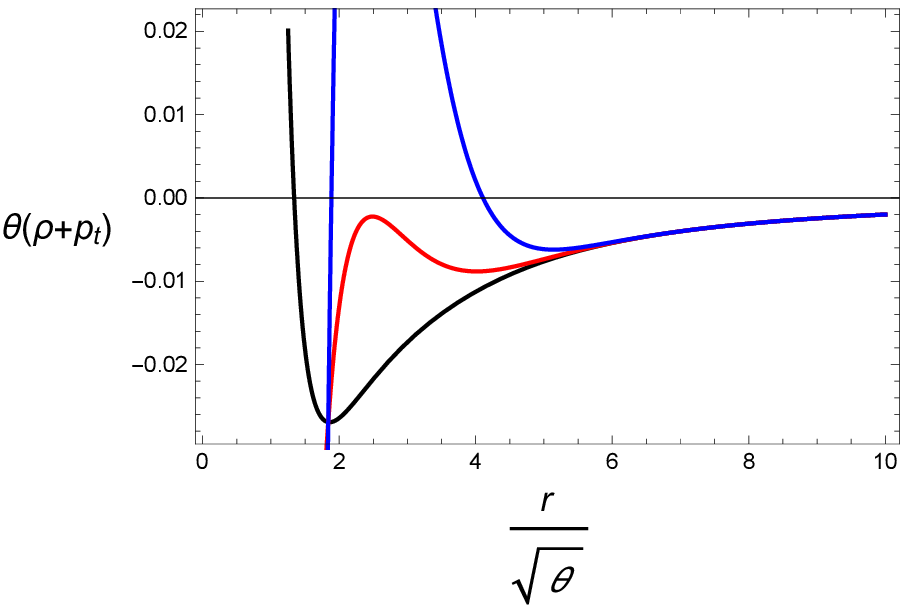, width=.45\linewidth,
height=1.4in}
\caption{\label{fig 4.3} This shows the development of $\theta(\rho+p_{r})$ and $\theta(\rho+p_{t})$ versus $\frac{r}{\sqrt{\theta}}$ for different values of $\frac{M}{\sqrt{\theta}}$ in the framework of Gaussian distribution.}
\end{figure}

Now we will discuss some interesting aspects of the obtained shape
function $b(r)$ which are considered as essential criteria for
wormholes existence. For this purpose, we choose some suitable
values of different free parameters involved. It is obvious that
Eq.(\ref{4.15}) depends on the coupling parameter $\lambda$, first
we need to fix this parameter in order to analyze the results more
comprehensively. Herein, we set $\lambda=2$ and represent the shape
function $b(r)$ of the form $b(\frac{r}{\sqrt{\theta}})$ which
depends on $\frac{M}{\sqrt{\theta}}$, dimensionless constant $C_2$
and integration constant $D_1$. One can choose $D_1=0$ as suggested
in previous studies \cite{rahamanPLB}, however in our case pick the
suitable value of $D_1$ depending on $\lambda$. For Gaussian
distributed non-commutative framework, we set $C_2=2$ and
$D_1=-2\sqrt[-5]{\theta}$. The throat of wormhole is located at
$\frac{r}{\sqrt{\theta}}=\frac{r_0}{\sqrt{\theta}}$, where
$b(\frac{r}{\sqrt{\theta}})=\frac{r_0}{\sqrt{\theta}}$. For
$\frac{M}{\sqrt{\theta}}=0.2$ (black curve), the throat of wormhole
is located at $\frac{r_0}{\sqrt{\theta}}=1.678$, whereas for the
other two values $\frac{M}{\sqrt{\theta}}=2$ (red curve) and
$\frac{M}{\sqrt{\theta}}=10$ (blue curve),
$b(\frac{r}{\sqrt{\theta}})-(\frac{r}{\sqrt{\theta}})$ crosses the
horizontal axis at $\frac{r_0}{\sqrt{\theta}}=2.563$ and
$\frac{r_0}{\sqrt{\theta}}=3.364$ respectively. It is noted that
position of the throat is increasing with the increase of smeared
mass distribution $M$ as shown in left plot of Figure \ref{fig 4.1}.
Right plot in Figure \ref{fig 4.1} shows that shape function has
increasing behavior for Gaussian distribution for different values
of $\frac{M}{\sqrt{\theta}}$. Validity of flaring out condition
$b^{'}(\frac{r}{\sqrt{\theta}})<1$ for
$\frac{r}{\sqrt{\theta}}>\frac{r_0}{\sqrt{\theta}}$ is evident from
left plot of Figure \ref{fig 4.2}. Right plot in Figure \ref{fig
4.2} indicates that
$\frac{b(\frac{r}{\sqrt{\theta}})}{\frac{r}{\sqrt{\theta}}}<1$ for
$(\frac{r}{\sqrt{\theta}})>(\frac{r_0}{\sqrt{\theta}})$, which is an
essential requirement for a shape function. We find that
$\frac{b(\frac{r}{\sqrt{\theta}})}{\frac{r}{\sqrt{\theta}}}\rightarrow4/5$
as $\frac{r}{\sqrt{\theta}}\rightarrow\infty$ as presented in Figure
\ref{fig 4.2}. In Figure \ref{fig 4.3}, we presented the graphical
behavior of the null energy conditions $\theta(\rho+p_r)$ and
$\theta(\rho+p_t)$. It can be seen that NEC is violated so that the
existence of wormholes requires exotic matter.

\subsection{Wormhole Existence with Lorentzian Distribution}

Here we discuss the case of noncommutative geometry with the
reference of Lorentzian distribution. In Lorentzian distribution, we
take density function given by Eq.(\ref{n2}). Comparing
Eqs.(\ref{n2}) and (\ref{4.7}), we get
\begin{equation}\label{4.16}
\frac{-2 C_2^2 (\lambda -1)+(6 \lambda -2) \psi ^2(r)+2 (3 \lambda -2) r \psi (r) \psi '(r)}{2 C_2^2 \left(2 \lambda ^2+1\right)
r^2}=\frac{M \sqrt{\theta}}{\pi^{2}(r^{2}+\theta)^{2}}.
\end{equation}
Solving this differential equation, we get the value $\psi^{2}(r)$
as follows
\begin{eqnarray}
\psi^{2}(r)&=&\frac{C_{2}^{2} r^{\frac{3 \lambda }{3 \lambda -2}+\frac{2-6 \lambda }{3 \lambda -2}+1}}{\pi ^2 \sqrt{\theta } (3 \lambda
-1)}\Bigg[\pi ^2 \sqrt{\theta } (\lambda -1)+\left(2 \lambda ^2 M+M\right)\times \, _2F_1\left(1,\frac{1-3
\lambda }{2-3 \lambda };\frac{3-6 \lambda }{2-3 \lambda };-\frac{r^2}{\theta }\right)-\left(2 \lambda ^2 M+M\right)
\nonumber\\&\times&\, _2F_1\left(2,\frac{1-3 \lambda }{2-3 \lambda };\frac{3-6 \lambda }{2-3 \lambda };-\frac{r^2}{\theta
}\right)\Bigg]+D_{2} r^{\frac{2-6 \lambda }{3 \lambda -2}},\label{4.17}
\end{eqnarray}
where $D_{2}$ is an integration constants and $_2F_1$ is a
hypergeometric function which is defined by
\begin{equation*}
_2F_1(a,b,c,z)=\sum^{\infty}_{n=0}\frac{(a)_{n}(b)_{n}}{c_{n}}\frac{t^{n}}{n!}.
\end{equation*}
Using this value of $\psi^{2}(r)$ in Eqs.(\ref{4.8}) and
(\ref{4.9}), we get the exact values of radial and tangential
pressures given by
\begin{eqnarray}
p_{r}&=&\frac{1}{\lambda +1}\Bigg[\frac{6 D_{2} (\lambda -1)
r^{\frac{2}{2-3 \lambda }-4}}{C_{2}^{2} (3 \lambda -2)}+\frac{6
(\lambda -1)\left(2 \lambda ^2+1\right) M \Gamma \left(\frac{1}{3
\lambda -2}\right)}{\pi ^2 \sqrt{\theta } (3 \lambda -2)^3
r^2}+\frac{\sqrt{\theta } \lambda  (\lambda +5) M}{\pi ^2 (3 \lambda
-2) \left(\theta +r^2\right)^2}-\frac{2 (\lambda +1)}{(3 \lambda -1)
r^2}\nonumber\\&\times&\left(\, _2\tilde{F}_1\left(1,1+\frac{1}{3
\lambda -2};2+\frac{1}{3 \lambda -2};-\frac{r^2}{\theta }\right)-\,
_2\tilde{F}_1\left(2,1+\frac{1}{3 \lambda
-2};2+\frac{1}{3 \lambda -2};-\frac{r^2}{\theta }\right)\right)\Bigg],\label{4.18}\\
p_{t}&=&\frac{1}{\pi ^2 (2-3 \lambda )^2 (\lambda +1) r^4}\Bigg[\frac{1}{C_{2}^{2} (3 \lambda -1) \left(\theta
+r^2\right)^2}\left(C_{2}^{2} r^2 \left(\sqrt{\theta } (3 \lambda -1) M \left(6
\theta  \lambda  \left(2 \lambda ^2+1\right)+(\lambda  (\lambda  (27 \lambda +5)+2)-4) r^2\right)
\right.\right.\nonumber\\&\-&\left.\left.\pi ^2 (2-3 \lambda )^2
(\lambda +1) \left(\theta +r^2\right)^2\right)-6 \pi ^2 D_{2} \lambda  (9
(\lambda -1) \lambda +2) \left(\theta +r^2\right)^2 r^{\frac{2}{2-3 \lambda }}\right)\nonumber\\&-&\frac{6 \lambda  \left(2
\lambda ^2+1\right) M r^2 \, _2F_1\left(1,1+\frac{1}{3 \lambda -2};2+\frac{1}{3 \lambda -2};-\frac{r^2}{\theta }\right)}{\sqrt{\theta
}}\Bigg],\label{4.19}
\end{eqnarray}
where $_2\tilde{F}_1$ is a regularized hypergeometric function which
is defined as
\begin{equation*}
_2\tilde{F}_1=\sum^{\infty}_{n=0}\frac{(a)_{n}(b)_{n}}{\Gamma c_{n}}\frac{t^{n}}{n!}.
\end{equation*}
One can calculate the shape function $b(r)$ for Lorentizian
distribution as follows
\begin{eqnarray}
b(r)&=&\frac{-r}{\pi ^2 C_2^2 \sqrt{\theta } (3 \lambda -1)
r}\left.\pi ^2 \sqrt{\theta } \left({C_2}^2 (\lambda -1) r+D_2(3
\lambda -1)r^{\frac{3 \lambda }{2-3 \lambda
}}\right)\right.\nonumber\\&+&\left.a^2 \left(2 \lambda ^2+1\right)
M r \left(\, _2F_1\left(1,1+\frac{1}{3 \lambda -2};2+\frac{1}{3
\lambda -2};-\frac{r^2}{\theta
}\right)\right.\right.\nonumber\\&-&\left.\left.\,
_2F_1\left(2,1+\frac{1}{3 \lambda -2};2+\frac{1}{3 \lambda
-2};-\frac{r^2}{\theta }\right)\right)\right)+r.\label{4.19}
\end{eqnarray}
\begin{figure}
\centering \epsfig{file=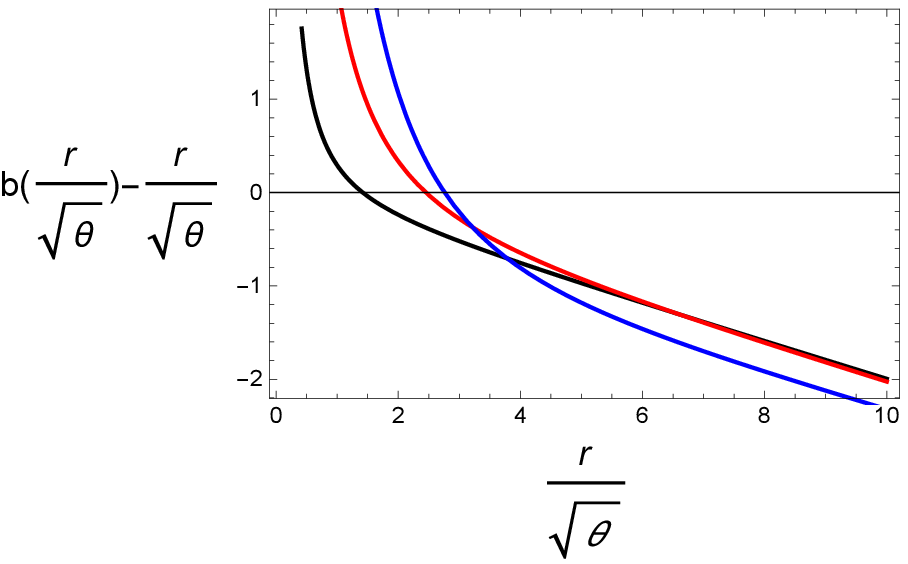, width=.45\linewidth,
height=1.4in}\epsfig{file=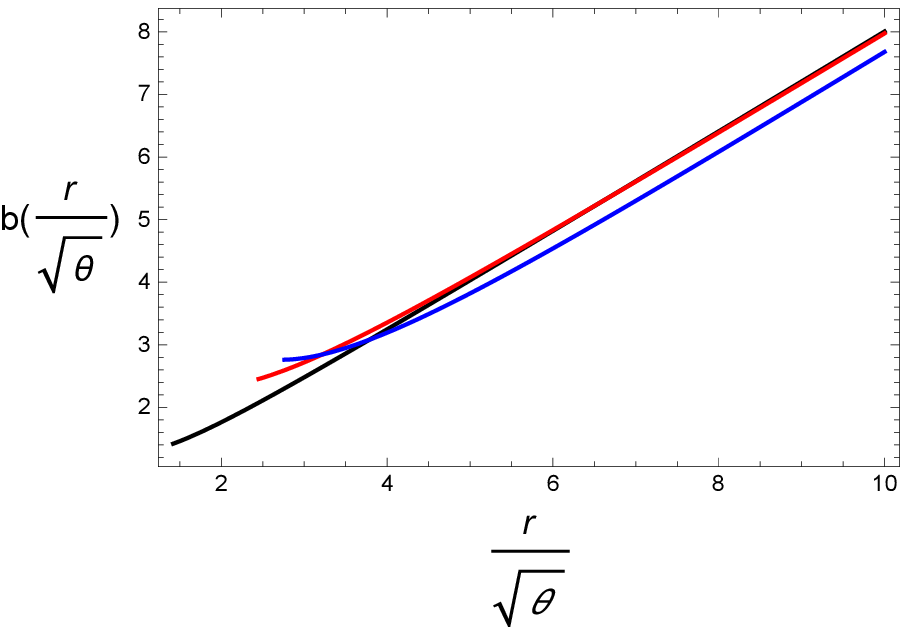, width=.45\linewidth,
height=1.4in} \caption{\label{fig 4.4} Evolution of
$b(\frac{r}{\sqrt{\theta}})-(\frac{r}{\sqrt{\theta}})$ and
$b(\frac{r}{\sqrt{\theta}})$ versus $\frac{r}{\sqrt{\theta}}$ for
different values of $\frac{M}{\sqrt{\theta}}$. Herein for Lorentzian
distribution, $\frac{M}{\sqrt{\theta}}=0.2$ represents the black
curve, $\frac{M}{\sqrt{\theta}}=2$ represents the red curve and
$\frac{M}{\sqrt{\theta}}=10$ represents the blue curve.}
\end{figure}
\begin{figure}
\centering \epsfig{file=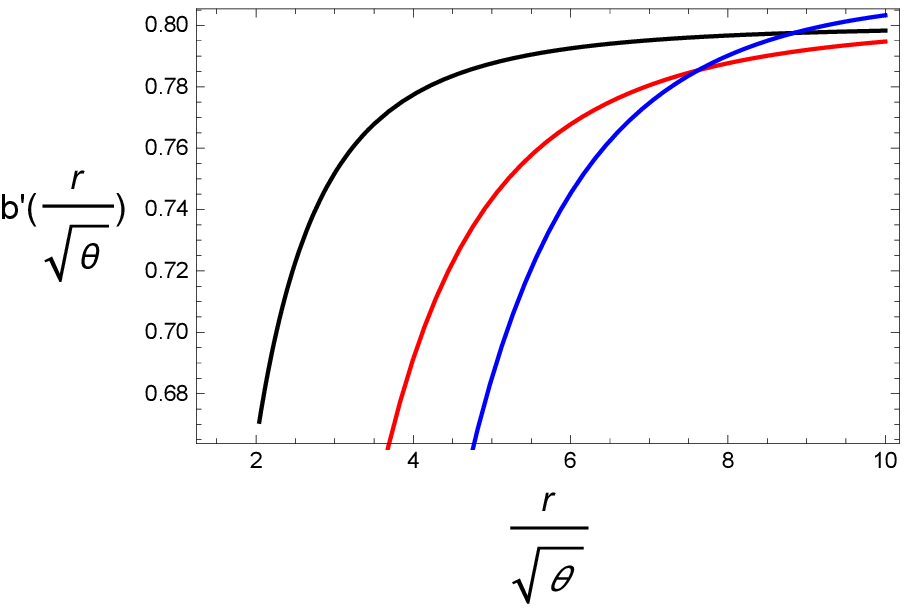, width=.45\linewidth,
height=1.4in}\epsfig{file=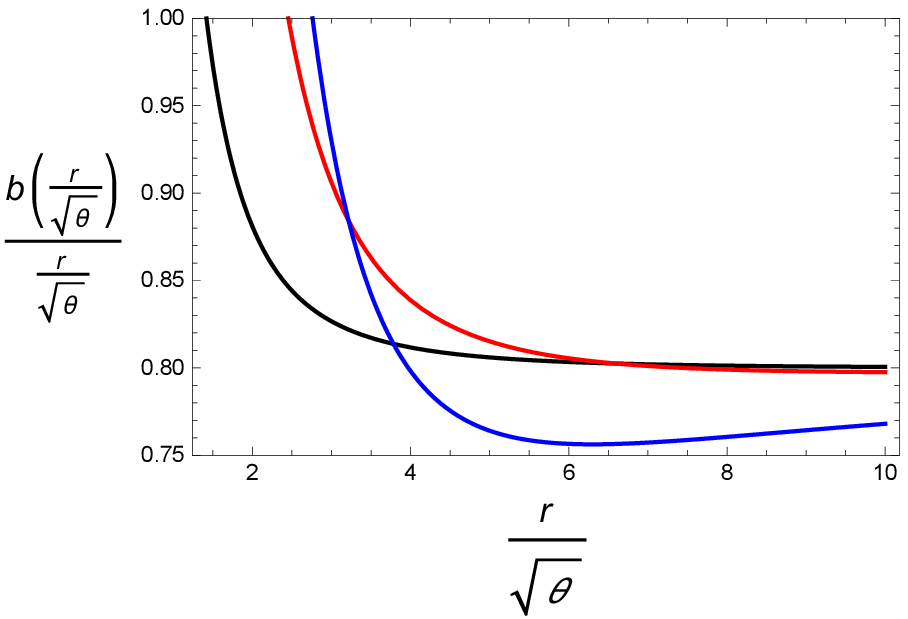, width=.45\linewidth,
height=1.4in} \caption{\label{fig 4.5} This indicates the behavior
of $b'(\frac{r}{\sqrt{\theta}})$ and
$\frac{b(\frac{r}{\sqrt{\theta}})}{\frac{r}{\sqrt{\theta}}}$ versus
$\frac{r}{\sqrt{\theta}}$ for different values of
$\frac{M}{\sqrt{\theta}}$ in the framework of Lorentzian
distribution.}
\end{figure}
\begin{figure}
\centering \epsfig{file=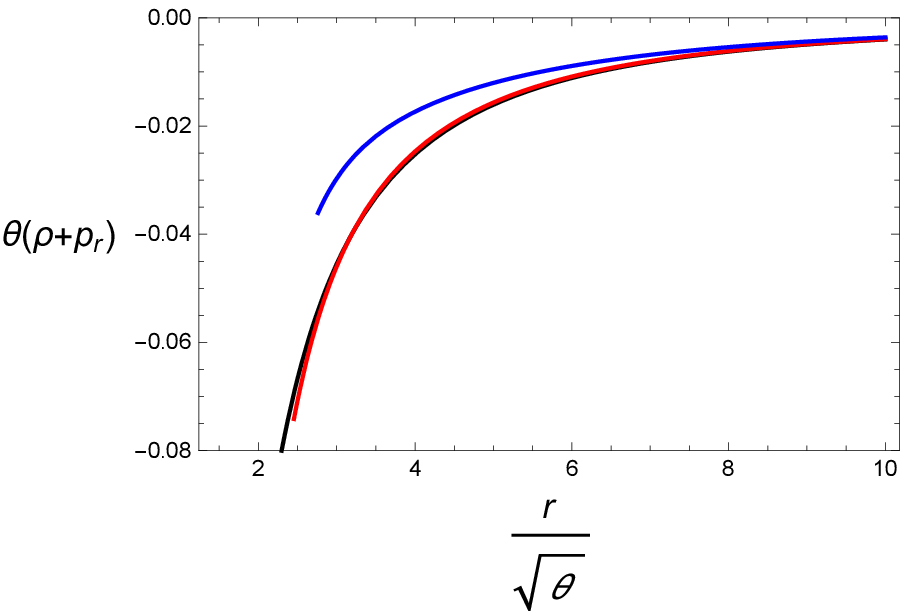, width=.45\linewidth,
height=1.4in}\epsfig{file=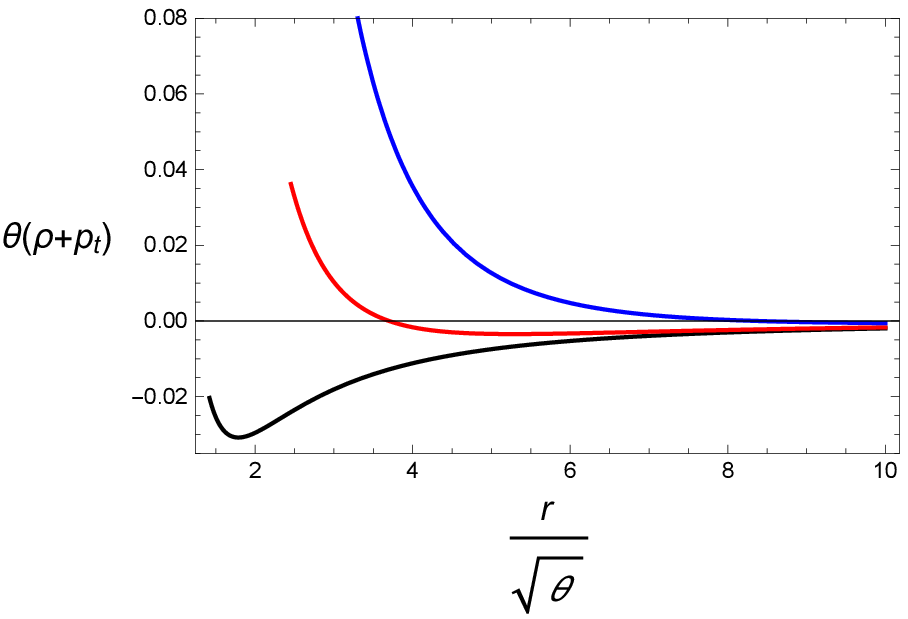, width=.45\linewidth,
height=1.4in} \caption{\label{fig 4.6} This shows the development of
$\theta(\rho+p_{r})$ and $\theta(\rho+p_{t})$ versus
$\frac{r}{\sqrt{\theta}}$ for different values of
$\frac{M}{\sqrt{\theta}}$ in the framework of Lorentzian
distribution.}
\end{figure}

Here, we discuss some properties of shape function $b(r)$ in the
background of non-commutative Lorentzian distribution. It can be
seen that Eq.(\ref{4.19}) depends on mass $M$, $\theta$ (the
non-commutative parameter) and the coupling parameter $\lambda$.
Initially, we select the particular value for the coupling parameter
$\lambda$ and analyze the results depending on the choice of other
parameters. Herein, we set $\lambda=2$ and represent the shape
function $b(r)$ of the form $b(\frac{r}{\sqrt{\theta}})$ which
depends on $\frac{M}{\sqrt{\theta}}$, dimensionless constant $C_2$
and integration constant $D_2$. $D_2$ can be selected as null
\cite{rahamanPLB}, however in this case, we pick the suitable value
of $D_2$ depending on the choice $\lambda$. For Lorentzian
distributed non-commutative framework, we set $C_2=2$. The throat of
wormhole is located at
$\frac{r}{\sqrt{\theta}}=\frac{r_0}{\sqrt{\theta}}$, where
$b(\frac{r}{\sqrt{\theta}})=\frac{r_0}{\sqrt{\theta}}$. We explore
the evolution of shape function depending on the choice of
$\frac{M}{\sqrt{\theta}}$. In left plot of Figure \textbf{4}, we
present the evolution of $b(\frac{r}{\sqrt{\theta}})$ versus
$\frac{r}{\sqrt{\theta}}$, it can be seen that for
$\frac{M}{\sqrt{\theta}}=0.2$ (black curve) with
$D_2=-2\sqrt[-5]{\theta}$, the throat of wormhole is located at
$\frac{r_0}{\sqrt{\theta}}=1.42$. For $\frac{M}{\sqrt{\theta}}=2$
(red curve) with $D_2=-10\sqrt[-5]{\theta}$, the location of throat
is at $\frac{r_0}{\sqrt{\theta}}=2.45$ whereas for
$\frac{M}{\sqrt{\theta}}=10$ (blue curve) with
$D_2=-25\sqrt[-5]{\theta}$,
$b(\frac{r}{\sqrt{\theta}})-(\frac{r}{\sqrt{\theta}})$ crosses the
horizontal axis at $\frac{r_0}{\sqrt{\theta}}=2.76$. It is deduced
that position of throat increases depending on the choice of smeared
mass distribution $M$ in similar fashion as in in Gaussian
distribution. We also present the evolution of
$b(\frac{r}{\sqrt{\theta}})$ on the right side of Figure \ref{fig
4.4} for different values of $\frac{M}{\sqrt{\theta}}$. Left plot of
Figure \ref{fig 4.5} presents the evolution of flaring out condition
which interprets $b^{'}(\frac{r}{\sqrt{\theta}})<1$ for
$\frac{r}{\sqrt{\theta}}>\frac{r_0}{\sqrt{\theta}}$ in all the
cases. We also evaluate
$\frac{b(\frac{r}{\sqrt{\theta}})}{\frac{r}{\sqrt{\theta}}}$ in the
limit of $\frac{r}{\sqrt{\theta}}\rightarrow\infty$, it is found
that
$\frac{b(\frac{r}{\sqrt{\theta}})}{\frac{r}{\sqrt{\theta}}}\rightarrow4/5$
similar to the previous Gaussian distribution case. The dynamical
behavior of null energy conditions $\theta(\rho+p_r)$ and
$\theta(\rho+p_t)$ is shown in Figure \ref{fig 4.6} and show similar
evolution as in Gaussian distribution.

\section{Equilibrium Condition}

In this segment, we explore the stability of obtained wormhole
solutions for both non-commutative distributions using equilibrium
condition. For this purpose, we take Tolman-Oppenheimer-Volkov
equation which is given by
\begin{equation}\label{3.27}
\frac{dp_{r}}{dr}+\frac{\sigma^{'}}{2}(\rho+p_{r})+\frac{2}{r}(p_{r}-p_{t})=0,
\end{equation}
where $\sigma(r)=2\Phi(r)$. This equation determines the equilibrium state of
configuration by taking the gravitational, hydrostatic as well as the
anisotropic forces (arising due to anisotropy of matter) into account. These
forces are defined by the following relations:
\begin{equation*}
F_{g}=-\frac{\sigma^{'}(\rho+p_{r})}{2},\;\;\;\;\;\;\;\;F_{h}=-\frac{dp_{r}}{dr},
\;\;\;\;\;\;\;\;F_{a}=2\frac{(p_{t}-p_{r})}{r},
\end{equation*}
and thus Eq.(\ref{3.27}) takes the form given by
\begin{equation}\label{3.28}
F_{a}+F_{g}+F_{h}=0.
\end{equation}
Firstly, we calculate these forces $F_{g}$, $F_{h}$ and $F_{a}$ for
Gaussian distribution as follows
\begin{eqnarray*}
F_{g}&=&-\frac{1}{8 \pi ^{3/2} \theta ^{3/2} r}\Bigg[ \frac{e^{-\frac{r^2}{4 \theta }}}{C_{2}^{2} \left(9 \lambda ^3-7 \lambda +2\right)
r^4}\bigg[ 2 C_{2}^{2} r^2 \bigg[(\lambda  (\lambda  (6 \lambda +7)-6)+1) M
r^2-8 \pi ^{3/2} \theta ^{3/2} \left(3 \lambda ^2+\lambda -2\right) e^{\frac{r^2}{4 \theta }}\bigg] \nonumber \\ &+&48 \pi ^{3/2} D_{1}
\theta ^{3/2}
(\lambda -1) (3 \lambda -1) e^{\frac{r^2}{4 \theta }} r^{\frac{2}{2-3 \lambda
}}\bigg]-\frac{6 (\lambda -1) \left(2 \lambda ^2+1\right) M E_{\frac{1}{2-3 \lambda }-1}\left(\frac{r^2}{4 \theta }\right)}{(2-3 \lambda
)^2 (\lambda +1)}\Bigg],\\
F_{h}&=&-\frac{1}{8 (\lambda +1)}\Bigg[ -\frac{288 D_{1} (\lambda -1) (2 \lambda -1) r^{\frac{2}{2-3 \lambda }-5}}{C_{2}^{2} (2-3
\lambda
)^2}+\frac{3 (\lambda -1) \left(2 \lambda ^2+1\right) M r E_{\frac{1}{2-3 \lambda }-2}\left(\frac{r^2}{4 \theta
}\right)}{\pi ^{3/2} \theta ^{5/2} (2-3 \lambda )^2}-\frac{\lambda  (\lambda +5) M r e^{-\frac{r^2}{4 \theta }}}{2 \pi ^{3/2} \theta
^{5/2} (3 \lambda -2)}\nonumber\\&+&\frac{16 (\lambda  (3 \lambda +2)+1)}{\left(2 \lambda ^2+1\right) r^3}+\frac{16 \left(-5
\lambda ^3+\lambda ^2+\lambda +3\right)}{(3 \lambda -1) \left(2 \lambda ^2+1\right) r^3}\Bigg],\\
F_{a}&=&\frac{1}{2 \pi ^{3/2} C_{2}^{2} \theta ^{3/2} (2-3 \lambda )^2 (\lambda +1) (3 \lambda -1) r^5}\Bigg[e^{-\frac{r^2}{4 \theta }}
\left((3 \lambda -2) \left(C_{2}^{2} r^2 \left((3 \lambda -1) \left(2 \lambda
^2+1\right) M r^2+4 \pi ^{3/2} \theta ^{3/2} (\lambda +1) \right.\right.\right.\nonumber\\ &\times&\left.\left.\left. (3 \lambda -2)
e^{\frac{r^2}{4 \theta
}}\right)-24 \pi ^{3/2} D_{1} \theta ^{3/2} (\lambda  (6 \lambda -5)+1)
e^{\frac{r^2}{4 \theta }} r^{\frac{2}{2-3 \lambda }}\right)+3 C_{2}^{2} (2 \lambda -1) (3 \lambda
-1) \left(2 \lambda ^2+1\right) M r^4 e^{\frac{r^2}{4 \theta }} E_{\frac{1}{2-3 \lambda
}-1}\left(\frac{r^2}{4 \theta }\right)\right)\Bigg].
\end{eqnarray*}
In case of Lorentizian distribution of non-commutative geometry,
these forces are given by
\begin{eqnarray*}
F_{g}&=&\frac{2}{r^5}\bigg[-\frac{3 D_{2} (\lambda -1) r^{\frac{2}{2-3 \lambda }}}{C_{2}^{2} \left(3 \lambda ^2+\lambda
-2\right)}+\frac{\sqrt{\theta } (1-\lambda  (2 \lambda +3)) M r^4}{\pi ^2 \left(3 \lambda ^2+\lambda -2\right) \left(\theta
+r^2\right)^2}+\frac{r^2}{3 \lambda -1}+\frac{3 (\lambda -1) \left(2 \lambda ^2+1\right) M r^2 \Gamma
\left(\frac{1}{3 \lambda -2}\right)}{\pi ^2 \sqrt{\theta } (\lambda +1) (3 \lambda -2)^3}\nonumber \\ &\times& \bigg[\,
_2\tilde{F}_1\left(2,1+\frac{1}{3
\lambda -2};2+\frac{1}{3 \lambda -2};-\frac{r^2}{\theta }\right)-\,
_2\tilde{F}_1\left(1,1+\frac{1}{3 \lambda -2};2+\frac{1}{3 \lambda -2};-\frac{r^2}{\theta }\right) \bigg]\bigg],\\
F_{h}&=&\frac{4}{r^5}\bigg[\frac{r^2}{1-3 \lambda }+\frac{9 D_{2} (\lambda -1) (2 \lambda -1) r^{\frac{2}{2-3 \lambda }}}{C_{2}^{2} (2-3
\lambda )^2 (\lambda +1)}-\frac{3 \sqrt{\theta } (\lambda -1) \left(2 \lambda ^2+1\right) M r^4}{\pi ^2 (2-3 \lambda )^2 (\lambda +1)
\left(\theta +r^2\right)^2}+\frac{\sqrt{\theta } \lambda  (\lambda +5) M r^6}{\pi ^2 \left(3 \lambda ^2+\lambda
-2\right) \left(\theta +r^2\right)^3}\nonumber\\&+&\frac{9 (\lambda -1) (2 \lambda -1) \left(2 \lambda ^2+1\right) M r^2 \Gamma
\left(\frac{1}{3
\lambda -2}\right)}{\pi ^2 \sqrt{\theta } (2-3 \lambda )^4 (\lambda +1)}\times\left(\,
_2\tilde{F}_1\left(1,1+\frac{1}{3 \lambda -2};2+\frac{1}{3 \lambda -2};-\frac{r^2}{\theta
}\right)\right.\nonumber\\&-&\left.\, _2\tilde{F}_1\left(2,1+\frac{1}{3 \lambda -2};2+\frac{1}{3 \lambda
-2};-\frac{r^2}{\theta }\right)\right)\bigg],\\
F_{a}&=&\frac{2}{\pi ^2 (2-3 \lambda )^2 (\lambda +1) r^5}\Bigg[\frac{1}{C_{2}^{2} (3 \lambda -1) \left(\theta
+r^2\right)^2}\bigg[C_{2}^{2} r^2 \bigg[2 \sqrt{\theta } (3 \lambda -1)\left(2
\lambda ^2+1\right) M \left(6 \theta  \lambda -3 \theta +(9 \lambda -5) r^2\right) \nonumber \\ &+&\pi ^2 (\lambda +1) (2-3 \lambda )^2
\left(\theta
+r^2\right)^2\bigg]-6 \pi ^2 D_{2} (2 \lambda -1) (3 \lambda -2) (3 \lambda -1) \left(\theta
+r^2\right)^2 r^{\frac{2}{2-3 \lambda }}\bigg]\nonumber\\&-&\frac{6 (2 \lambda -1) \left(2 \lambda ^2+1\right) M r^2 \,
_2F_1\left(1,1+\frac{1}{3 \lambda -2};2+\frac{1}{3 \lambda -2};-\frac{r^2}{\theta }\right)}{\sqrt{\theta }}\Bigg].
\end{eqnarray*}
The graphical illustration of these forces is given in Figure
\textbf{7}. The left graph indicates the behavior of these forces
for Gaussian distribution, while the right graph corresponds to
Lorentzian distribution. It is evident from these graphs that the
stability of configuration has been attained as gravitational and
anisotropic forces show opposite behavior to hydrostatic force and
hence cancel each other's effect.
\begin{figure}
\centering \epsfig{file=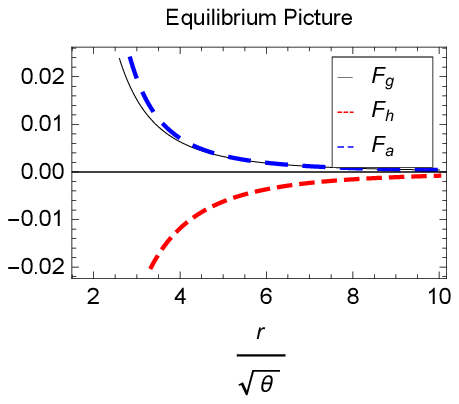, width=.45\linewidth,
height=1.4in}\epsfig{file=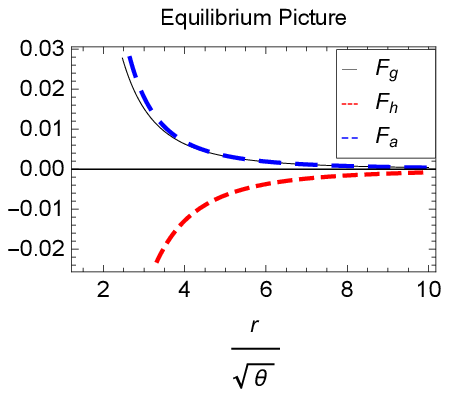, width=.45\linewidth,
height=1.4in} \caption{\label{fig 4.7} This shows the graphical
illustration of $F_{a}$, $F_{g}$ and $F_{h}$ forces versus $r$ for
Gaussian and Lorentizian distribution, in the left and right panels,
respectively. In left plot we set $\frac{M}{\sqrt{\theta}}=0.2$ and
$\frac{r_0}{\sqrt{\theta}}=1.678$, whereas in right plot the
selected parameters are $\frac{M}{\sqrt{\theta}}=0.2$ and
$\frac{r_0}{\sqrt{\theta}}=1.420$.}
\end{figure}

\section{Active Gravitational Mass}

The active gravitational mass within the region from the throat
$r_0$ up to the radius $R$ can be found by using the relation
$M_{active}=4 \pi  \int_{\text{r0}}^R \rho  r^2 \, dr$. For Gaussian
distribution, it is given by
\begin{eqnarray}\label{4.21}
M_{active}=M \left[\text{erf}\left(\frac{r}{2 \sqrt{\theta }}\right)-\frac{r e^{-\frac{r^2}{4 \theta }}}{\sqrt{\pi } \sqrt{\theta
}}\right]^{R}_{r_{0}}.
\end{eqnarray}
In case of Lorentizian distribution, we have
\begin{eqnarray}\label{4.22}
M_{active}=\frac{2 M}{\pi } \left[\tan ^{-1}\left(\frac{r}{\sqrt{\theta }}\right)-\frac{\sqrt{\theta } r}{\theta
+r^2}\right]^{R}_{r_{0}}.
\end{eqnarray}
We observe that by above equation the active gravitational mass $M_{active}$ of the wormhole is positive under the constraint
$\text{erf}\left(\frac{r}{2 \sqrt{\theta }}\right)>\frac{r e^{-\frac{r^2}{4 \theta }}}{\sqrt{\pi } \sqrt{\theta }}$ for Gaussian
distribuion and $\tan ^{-1}\left(\frac{r}{\sqrt{\theta }}\right)>\frac{\sqrt{\theta } r}{\theta +r^2}$ for Lorentizian distribution.
The physical nature of active gravitational mass can be seen from Figure \textbf{8}.
\begin{figure}
\centering \epsfig{file=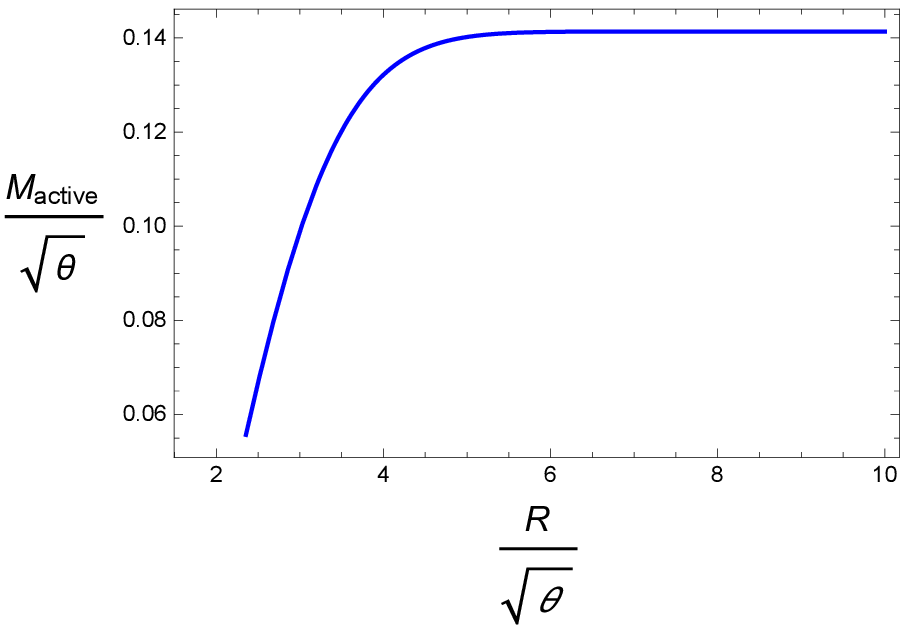, width=.45\linewidth,
height=1.4in}\epsfig{file=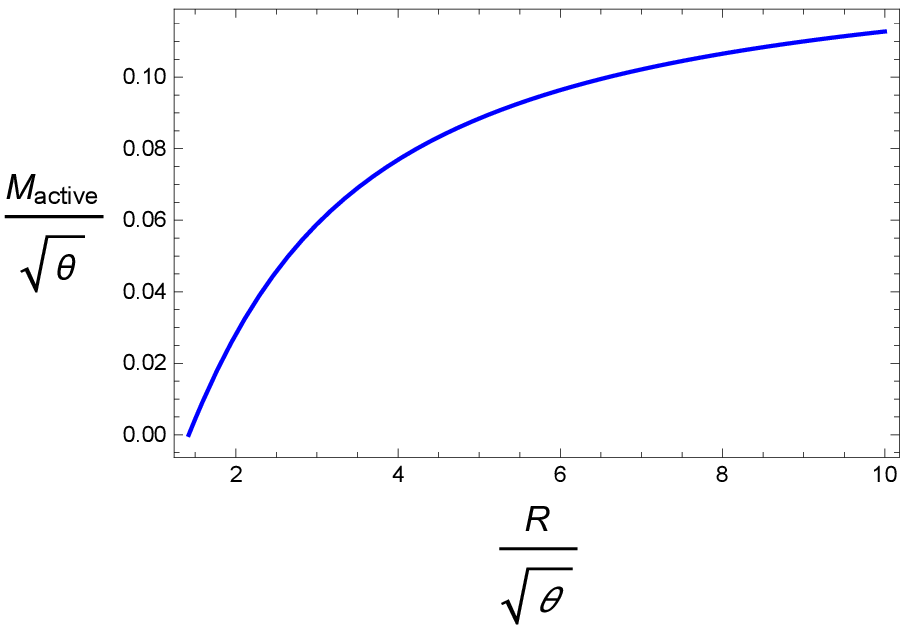, width=.45\linewidth,
height=1.4in} \caption{\label{fig 4.8} This shows the graphical
illustration of $\frac{M_{active}}{\sqrt{\theta}}$ versus
$\frac{R}{\sqrt{\theta}}$ for Gaussian and Lorentizian
distributions, in the left and right panels, respectively. In left
plot we set $\frac{M}{\sqrt{\theta}}=0.2$ and
$\frac{r_0}{\sqrt{\theta}}=1.678$, whereas in right plot the
selected parameters are $\frac{M}{\sqrt{\theta}}=0.2$ and
$\frac{r_0}{\sqrt{\theta}}=1.420$.}
\end{figure}

\section{Conclusion}

In the present manuscript, we have explored the existence of
spherically symmetric wormhole solutions by taking corresponding
CKVs into account. For this purpose, we consider anisotropic matter
contents along with Gaussian and Lorentzian distributions of
non-commutative geometry. The concept of introducing CKVs and
non-commutative distributions for finding solutions is not a new
approach. These concepts have already been used in literature in
various contexts like in investigating the existence of wormholes
and black holes in different gravitational theories. However, the
present work provides a unique discussion in the sense that such
approach of conformal motion along with non-commutative distribution
has not been used before in $f(R,T)$ theory. In this respect,
Kuhfittig \cite{1*} has used non-commutative geometry in $f(R)$
gravity to discuss different forms of $f(R)$ function by taking
different choices of shape functions into account. In
\cite{2*}, the same author introduced CKVs to check the existence
and stability of wormholes by using phantom energy, i.e.,
$p=\omega\rho;~\omega<-1$. Jamil et al. \cite{3*}
discussed the wormhole solutions by considering non-commutative
geometry in $f(R)$ gravity. In another study \cite{4*},
non-commutative distributions are used to reconstruct $f(R)$ models
by taking two choices of shape function. In \cite{5*}, Singh et al. have introduced the aspects of non-commutative geometry to
discuss the rotating black string where they also examined the
stability by checking thermodynamical properties of solutions. In
another study, Ghosh et al. \cite{Ghosh} introduced non-commutative
geometry to discuss the existence of Einstein-Gauss-Bonnet black
hole. Nicolini et al. \cite{6**} investigated the behavior of a noncommutative radiating Schwarzschild black hole. In our recent paper \cite{10}, we have also introduced the
concept of non-commutativity in $f(R,T)$ gravity (without including
CKVs) to check the possible existence and stability of wormhole
solutions. It is seen that due to complex form of resulting field
equations, we discussed wormhole solutions numerically there except
for few cases.

In the present paper, for both non-commutative distributions, the
use of CKVs simplifies the resulting field equations and
consequently leads to the analytical solutions of field equations
which meet all the necessary criteria for wormhole existence. We
have shown these properties of wormholes graphically. We have
firstly defined the possible CKVs of a general static spherically
symmetric spacetime which leads to simplicity in calculations
further. With the help of these CKVs, we have formulated a
relatively simplified form of field equations. By considering the
density functions of Gaussian and Lorentzian distributions of
non-commutative geometry, it is seen that the conformal killing
vector $\psi(r)$ is obtained in terms of exponential and
hypergeometric functions of mathematics. By using these CKVs, we
have explored the corresponding forms of shape functions and their
behavior graphically as shown in Figures \textbf{1-8}. By fixing the
arbitrary constants, graphical analysis has been done in terms of
dimension less variable, i.e., we plotted all the properties of
shape function versus $\frac{r}{\sqrt{\theta}}$ for three different
choices of $\frac{M}{\sqrt{\theta}}$. The obtained results can be
summarized as follows:
\begin{itemize}
\item the obtained shape functions, in both cases, are increasing
functions in nature versus $\frac{r}{\sqrt{\theta}}$.

\item the validity of flaring out property has been achieved in both
cases.

\item for the constructed shape functions, in both cases, $\frac{b(\frac{r}{\sqrt{\theta}})}{\frac{r}{\sqrt{\theta}}}$
approaches to a constant value of $\frac{4}{5}$ as
$\frac{r}{\sqrt{\theta}}\rightarrow\infty$.

\item it is seen that the values of wormhole throat increase as the values of
$\frac{M}{\sqrt{\theta}}$ increase in both Gaussian and Lorentzian
distributions.

\item At these wormhole throats, it has been shown graphically that
the condition $b'(\frac{r_0}{\sqrt{\theta}})<1$ holds in both cases.

\item the NEC bounds are violated and thus confirming the presence of exotic
matter that is a basic requirement for wormhole existence.

\item using Tolman-Oppenheimer-Volkov equilibrium condition, it is
seen from the graphical behavior of gravitational, hydrostatic and
anisotropic forces, the obtained wormhole solutions are stable.
Basically these forces balance each other's effect and hence leave a
stable configuration.

\item the forms of active gravitational mass show positive increasing behavior under
some certain constraints imposed on the free parameters in both
cases.
\end{itemize}

It would be interesting to explore the existence of wormhole
solutions using CKVs in other modified gravity theories.

\vspace{.5cm}

\section*{Acknowledgments}

``M. Zubair thanks the Higher Education Commission, Islamabad, Pakistan for its
financial support under the NRPU project with grant number
$\text{5329/Federal/NRPU/R\&D/HEC/2016}$''.

\end{document}